# Analytic Modeling of Starshades

Webster Cash

*Center for Astrophysics and Space Astronomy,*

*University of Colorado, Boulder, CO 80309, USA*

ABSTRACT

External occulters, otherwise known as starshades, have been proposed as a solution to one of the highest priority yet technically vexing problems facing astrophysics - the direct imaging and characterization of terrestrial planets around other stars. New apodization functions, developed over the past few years, now enable starshades of just a few tens of meters diameter to occult central stars so efficiently that the orbiting exoplanets can be revealed and other high contrast imaging challenges addressed. In this paper an analytic approach to analysis of these apodization functions is presented. It is used to develop a tolerance analysis suitable for use in designing practical starshades. The results provide a mathematical basis for understanding starshades and a quantitative approach to setting tolerances.

Keywords: coronagraphs, exoplanets, diffraction

# I. Introduction

Nearly everybody wants to know if Earth-like planets abound in the Universe. Are warm, watery paradises common, and does life arise everywhere it is given a chance? To answer these age-old questions requires a very good telescope capable of pulling the signal from a faint Earth-like planet out of the glare of its parent star. It will probably be necessary to look out to distances of 10 parsecs or more to have a good chance of finding such an Earth twin (Turnbull et al, 2011). But at that distance, the Earth is only thirtieth magnitude and hovers less than 0.1 arcseconds from the star.

This is a daunting challenge for telescope builders. An m=30 object, at 0.1 arcsecond angular separation, is at both the sensitivity limit and angular resolution limit of the Hubble Space Telescope. So an Earth-searching telescope has to be expensive and high quality if it is to be able to resolve and study the planetary system - even if there is no glare from the star.

The Terrestrial Planet Finder program encapsulated NASA's response. Two approaches were developed to building telescopes that could null out the parent star and thereby enable direct observation of the Habitable Zone. One approach uses high precision nulling between spacecraft in the mid-infrared to suppress the stellar glare (see for example Lawson et al, 2006). The other uses wavefront control and correction in an internal coronagraph to remove the central starlight (e.g. Guyon, et. al., 2006). Both approaches have proven to be difficult and expensive.

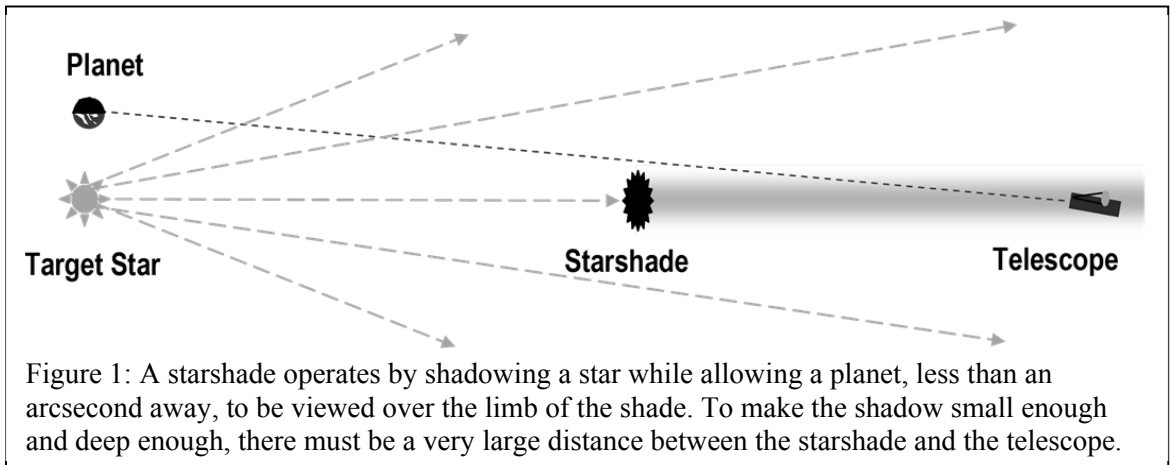

Figure 1: A starshade operates by shadowing a star while allowing a planet, less than an arcsecond away, to be viewed over the limb of the shade. To make the shadow small enough and deep enough, there must be a very large distance between the starshade and the telescope.

More recently, the idea of an external occulter (Spitzer, 1962) has been resurrected. The idea (shown schematically in Figure 1) is to keep the starlight from ever entering the telescope where it causes such havoc. A properly shaped device flown on a separate spacecraft can be moved into the line of sight such that it blots out the star. If this external occulter (which is often called a starshade) subtends a sufficiently small angle on the sky, it can blot out the star without impeding the light from the nearby planet. But this forces the shade onto a separate spacecraft. Even if the shade is only slightly larger than the telescope, it must be flown thousands of kilometers from the telescope in order to appear small enough.

However, diffraction around the starshade and into the telescope can be severe. This forces the starshade to be even larger and farther away. In 1985 Marchal presented the first serious diffraction analysis for external occulters. He showed that apodization functions could greatly reduce the size of an occulter compared to that required for a simple circular mask. He also suggested the use of petals to approximate a circularly symmetric function and thereby sidestep the problem of scattering through partially transmitting screens. But the size scales required to view Earth-like planets remained impractically high – occulters would have to be about a kilometer in diameter and fly at a million kilometers of separation. Copi and Starkman revisited this problem of suppression in 2000 and proposed a practical design that could suppress to the 4x10$^{-5}$ level.

A few years later it was shown there existed an apodization function that allows one to reduce the required diameter of an external occulter by over an order of magnitude (Cash 2006). The reduction of required diameter to a few tens of meters for the first time brought starshades into a size range that could be seriously considered for flight. The new function was the "offset hypergaussian" given by:

$$A(\rho) = 0 \qquad \text{for } \rho < a \qquad \qquad 1$$

and

$$A(\rho) = 1 - e^{-\left(\frac{\rho-a}{b}\right)^n} \qquad \text{for } \rho > a \qquad \qquad 2$$

In that paper it was shown how this new apodization function led to mission designs that would be capable of finding Earths and searching for life, yet appeared to be within the capability of current aerospace engineering techniques and space agency budget constraints. A generalized computer search by Vanderbei, Cady and Kasdin (2007) showed that the optimal apodization function strongly resembles an offset hypergaussian, and that diameter reductions of no more than about another 25% can be expected. This was not unexpected, because the offset hypergaussian already allows one to operate at only six Fresnel zones of radius. In section V of this paper the origins and tradeoffs between the computer optimized solution and the hypergaussian are addressed.

A great deal more work has transpired in studying these systems since. In particular, starshades are now embodied in space astronomy mission concepts called the New Worlds Observer (NWO; Cash et al, 2009) and THEIA (Kasdin et al, 2009). NWO nominally has design parameters of a=b=12.5m and n=6. This means that the shade is 62m across, from tip to tip. The diameter to the inflection point (2*(a+b)), which is more representative of the point at which the obscuration ends and the transmission of exoplanet light begins, is 50m. The New Worlds starshade flies at a nominal distance (F) of 80,000km from its telescope. At that distance the 25m radius to which exoplanets can be seen subtends 0.064 arcseconds, which is a small enough Inner Working Angle to allow observation of Earth-like planets at 10pc. It operates in the visible band from 0.3µ to 1µ wavelength. These baseline parameters are used throughout the paper when a nominal design is needed.

The search for the solution to the high-contrast occulter must be carried out with the full complexity of the Fresnel regime. A Fraunhoffer solution implies that, to good approximation, all the rays impinge upon the mask with the same phase. But an occulting mask fundamentally cannot operate in that manner. A shadow is formed only when the sum of electric fields outside the mask is small, thereby requiring a range of phases that sums to zero. A Fraunhoffer solution would require the mask to be restricted to a single zone and the sum of phases cannot be zero. So, to achieve a net zero electric field in the focal plane, the integral must extend out of the central zone at least into the first negative Fresnel half zone.

While it is quite remarkable that shadows of such extreme depth can be generated across just a few zones, that fact alone is not enough to justify their choice for use in the pursuit of exoplanets. First starshades must be understood so as to develop certainty that they are applicable in a practical and affordable manner. Unfortunately there is no long history of use that has created a body of generally accepted knowledge and analysis must start anew.

In addition to the analytic analysis discussed herein, practical demonstrations of small starshades have been performed in the laboratory.

Table I
List of Variables in Fresnel-Kirchoff Derivations

| | |
|---|---|
| **A** | E field amplitude |
| **S** | surface of integration |
| U | resultant electric field |
| P | point in shadow plane |
| $P_0$ | point of E field origination |
| r | distance $P_0$ to point in plane of integration |
| $r_0$ | distance $P_0$ to origin in plane of diffraction |
| r' | distance $P_0$ to point in z=0 plane |
| $r_1$ | $P_0$ height above z=0 plane |
| s | distance P to point in plane of integration |
| $s_0$ | distance P to origin in plane of intergration |
| s' | distance P to point in z=0 plane |
| $s_1$ | P height below z=0 plane |
| x,y,z | coordinates of diffraction plane |
| θ | angle between $P_0$ direction and normal to plane |
| **n** | normal vector to diffraction plane |
| λ | wavelength of light |
| λ' | λcosθ |
| k | 2π/λ |
| k' | 2π/λ' |

Scale models have now achieved shadows of depth sufficient to support observations of exoplanets (Schindhelm, 2007; Leviton 2007). So the basic performance of the apodization function has already been demonstrated.

There are two aspects to the modeling that are necessary for full understanding. First, the shadows need to be modeled analytically. Direct use of the equations of diffraction as applied to the apodization functions can give basic insight into the performance of the shades. Simple scaling laws and an understanding of the linkages between parameters can best be understood from such results.

Second, detailed computer modeling is needed. Just as raytracing is necessary for full understanding of the behavior and tolerancing of complicated geometrical optics systems, so too is full-up numerical modeling necessary to the design of starshades. This paper addresses both these needs.

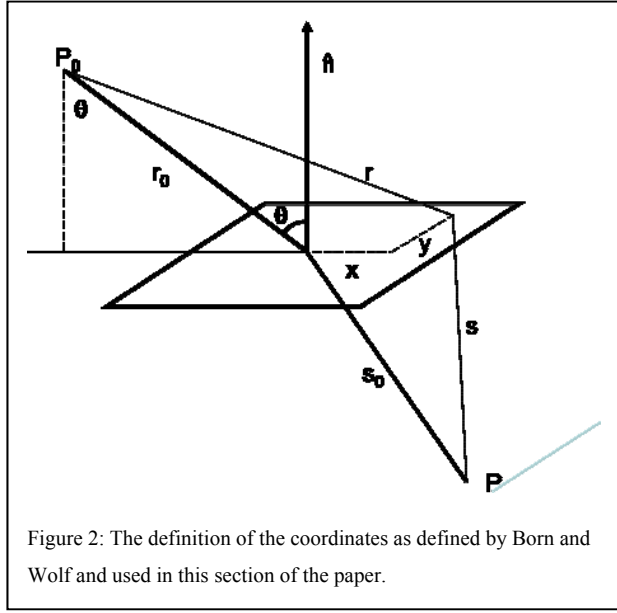

Figure 2: The definition of the coordinates as defined by Born and Wolf and used in this section of the paper.

## II. Analytic Analysis of the Problem

The analysis begins with some general discussion of the mathematics and physics that are needed to model the behavior of starshades. Since the goal is to reach accuracies below $10^{-12}$ in diffraction suppression, care must be exercised about the assumptions and approximations. As such, the analysis must begin with the most basic of electromagnetic equations and be systematically derived from there.

### A. Fresnel-Kirchoff Formulation

The starting point for the discussion will be the Fresnel-Kirchoff formula as presented by Born and Wolf (1999). We utilize their notation for the initial analysis (equations 3 through 21) up through the proof of the Fresnel approximation as summarized in Table I. The Fresnel-Kirchoff formula assumes that edge effects on the diffracting element are small, which will surely be the case with a large diffracting element like a starshade. There is some possibility of small effects near the tips and near the base of the petals of a starshade, so these effects will eventually have to be measured in the laboratory. But there is no reason to suspect that they will be significant.

The electric field U due to the radiation at point P is given by

$$U(P) = -\frac{iA}{2\lambda} \iint_S \frac{e^{ik(r+s)}}{rs} [\cos(n,r) - \cos(n,s)] dS \qquad 3$$

where r is the distance from the source to the surface and s is the distance from surface to P. A is the amplitude of the disturbance at unit distance from the source, $\lambda$ is the wavelength, k is $2\pi/\lambda$, cos(n,r) is the cosine between the local normal to the surface and the line from the source to that point, and the integration proceeds over the surface S.

Next, constrain S to be the z=0 plane, and define $r_0$ and $s_0$ to be the distances from the source ($P_0$) to the origin and the origin to P respectively as in Figure 2. Then, defining $r'$ and $s'$ to be

$$r' = \sqrt{r_0^2 + x^2 + y^2} \quad \text{and} \quad s' = \sqrt{s_0^2 + x^2 + y^2} \qquad 4$$

it is found that

$$U(P) = -\frac{iA}{2\lambda} \iint_S \frac{e^{ik(r'+s')}}{r's'} \left(\frac{r_0}{r'} + \frac{s_0}{s'}\right) dS \qquad 5$$

for the case where the plane of integration is perpendicular to the r-s line.

However, it is useful to generalize to the case where the plane of integration is tilted at an angle $\theta$ to the source to $P_0$-P line. In which case

$$r^2 = r_0^2 \cos^2\theta + (r_0 \sin\theta + x)^2 + y^2 = r'^2 + 2xr_0 \sin\theta \quad \text{and} \qquad 6$$

$$s^2 = s_0^2 \cos^2\theta + (s_0 \sin\theta - x)^2 + y^2 = s'^2 + 2xs_0 \sin\theta$$

which leads to

$$U(P) = -\frac{iA}{2\lambda}\cos\theta \iint_S \frac{e^{ik(r+s)}}{rs}\left(\frac{r_0}{r} + \frac{s_0}{s}\right)dxdy \qquad 7$$

The cosine term is the result of the oblique angle of the disturbance on the mathematical plane and must be accounted for.

### B. Babinet's Principle

To evaluate the diffraction into the shadow of a starshade integration must be carried out over the infinite plane outside of the obscuring mask. However, this tends to be impractical, so use Babinet's Principle allows the integration to proceed over the occulter only. Born and Wolf present the principle as

$$U = U_{mask} + U_{aperture} \qquad 8$$

which appears simple enough, but must be carefully applied. $U_{mask}$ is the equation 7 integral over that part of the plane that is opaque, while $U_{aperture}$ is the integral over the rest of the infinite plane. This equation is deceptively simple, and care must be taken with its use. When U is defined by Equation 7, its functional form can vary depending on the how the integral is set up. In particular, if the plane of integration is tilted (e.g. the starshade tilts out of alignment) then the value of U can be changed. This is an oddity of the Fresnel-Kirchoff formula, but must be included to avoid serious mathematical error in the application of Babinet's Principle.

Take the case of a line from $P_0$ to P running through the origin of the plane of integration, which lies $r_0$ from $P_0$ and $s_0$ from P. The disturbance at P will then be given by

$$U(P) = \frac{Ae^{ik(r_0 + s_0)}}{r_0 + s_0} \qquad 9$$

But evaluation of equations 6 and 7 gives a somewhat different answer.

In equation 7 make the substitutions

$$x' = \frac{x}{\cos\theta}, \; y' = \frac{y}{\cos\theta}, \; \lambda' = \lambda\cos\theta, \; r_0 = \frac{r_1}{\cos\theta}, \; s_0 = \frac{s_1}{\cos\theta} \qquad 10$$

To find that

$$U(P) = -\frac{iA}{2\lambda} \cos^2 \theta \iint_S \frac{e^{ik'(r'+s')}}{r's'} \left( \frac{r_1}{r'} + \frac{s_1}{s'} \right) dxdy \qquad 11$$

where

$$r'^2 = r_1^2 \cos^2 \theta + (r_1 \sin \theta + x)^2 + y^2 \quad \text{and} \qquad 12$$

$$s'^2 = s_1^2 \cos^2 \theta + (s_1 \sin \theta - x)^2 + y^2$$

But inspection of equation 11 shows that it must be identical to

$$U(P) = \cos^2 \theta \frac{A e^{ik'(r_1+s_1)}}{r_1 + s_1} \qquad 13$$

which is the same as

$$U(P) = \cos \theta \frac{A e^{ik(r_0+s_0)}}{r_0 + s_0} \qquad 14$$

Equation 14 yields a disturbance that differs from equation 9 by a factor of cosθ although it differs only in the definition of the plane over which the integration was performed, which should not affect the value of the disturbance, but appears to anyway. So whenever one sets up a calculation that has either $P_0$ or P off center, this mathematical artifact must be remembered.

### C. The Fresnel Approximation

For the case of a starshade, both $r_0$ and $s_0$ are very much larger than the size of the occulter that is to be integrated over. This allows use of the approximation first used by Fresnel. Start with equation 7 and recognize from equation 6 that

$$\frac{r}{r_0} = \sqrt{1 + \frac{x^2}{r_0^2} + \frac{y^2}{r_0^2} + \frac{2x \sin \theta}{r_0}} \approx 1 + \frac{x^2}{2r_0^2} + \frac{y^2}{2r_0^2} + \frac{x \sin \theta}{r_0} - \frac{x^4}{8r_0^4} - \frac{y^4}{8r_0^4} - \frac{x^2 \sin^2 \theta}{2r_0^2} + \ldots \qquad 15$$

$$\frac{s}{s_0} = \sqrt{1 + \frac{x^2}{s_0^2} + \frac{y^2}{s_0^2} - \frac{2x \sin \theta}{s_0}} \approx 1 + \frac{x^2}{2s_0^2} + \frac{y^2}{2s_0^2} - \frac{x \sin \theta}{s_0} - \frac{x^4}{8s_0^4} - \frac{y^4}{8s_0^4} - \frac{x^2 \sin^2 \theta}{2s_0^2} + \ldots$$

and that the terms in $x^4$ and $y^4$ are exceedingly small and may be safely dropped. Then

$$r + s \approx r_0 + s_0 + \frac{x^2}{2r_0} + \frac{y^2}{2r_0} + \frac{x^2}{2s_0} + \frac{y^2}{2s_0} - \frac{x^2 \sin^2 \theta}{2r_0} + \frac{x^2 \sin^2 \theta}{2s_0} \quad \text{and}$$

$$rs = r_0 s_0 \left[1 + \frac{x^2}{2r_0^2} + \frac{y^2}{2r_0^2} + \frac{x\sin\theta}{r_0} - \frac{x^2 \sin^2\theta}{2r_0^2}\right]\left[1 + \frac{x^2}{2s_0^2} + \frac{y^2}{2s_0^2} - \frac{x\sin\theta}{s_0} - \frac{x^2 \sin^2\theta}{2s_0^2}\right] \quad 16$$

Because $r_0$ is much greater than $s_0$, all terms with $r_0$ in the denominator may be dropped. Finally, all the terms in the product may be dropped, because the largest is $x\sin\theta/s_0$, which is of order $10^{-7}$. This results in:

$$r + s \approx r_0 + s_0 + \frac{x^2}{2s_0} + \frac{y^2}{2s_0} + \frac{x^2 \sin^2\theta}{2s_0} \quad \text{and} \quad rs = r_0 s_0 \quad 17$$

which may be substituted into equation 7 to find

$$U(P) = -\frac{iA}{\lambda}\cos\theta \frac{e^{ik(r_0+s_0)}}{r_0 s_0} \int e^{\frac{iky^2}{2s_0}} dy \int e^{\frac{ikx^2}{2s_0}} e^{-\frac{ikx^2 \sin^2\theta}{2s_0}} dx \quad 18$$

which becomes, when the plane is perpendicular to $P_0$-$P$,

$$U(P) = -\frac{iA}{\lambda}\frac{e^{ik(r_0+s_0)}}{r_0 s_0} \int e^{\frac{iky^2}{2s_0}} dy \int e^{\frac{ikx^2}{2s_0}} dx \quad 19$$

which is the usual form of the Fresnel approximation.

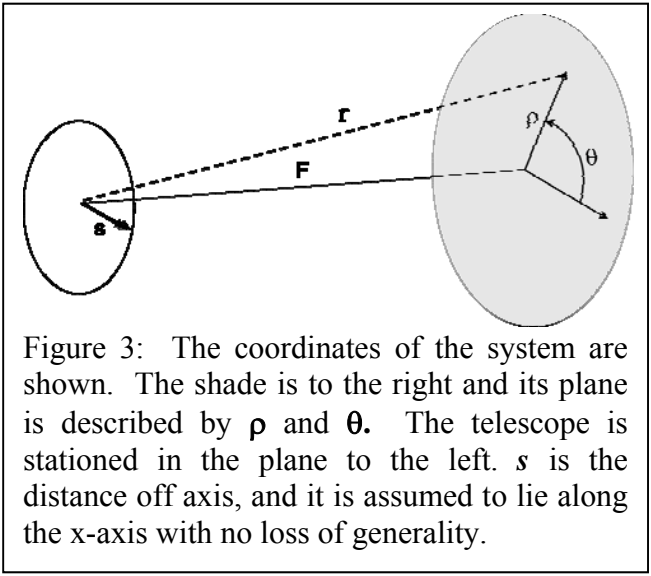

Figure 3: The coordinates of the system are shown. The shade is to the right and its plane is described by $\rho$ and $\theta$. The telescope is stationed in the plane to the left. $s$ is the distance off axis, and it is assumed to lie along the x-axis with no loss of generality.

Often, when one wishes to evaluate the shadow from a tilted starshade, the tilted aperture is approximated with its projection into the untilted plane, which simply means that in the x direction is integrated from $a*\cos\theta$ to $b*\cos\theta$ instead of a to b:

$$U(P) = -\frac{iA}{\lambda}\frac{e^{ik(r_0+s_0)}}{r_0 s_0} \int e^{\frac{iky^2}{2s_0}} dy \int_{a\cos\theta}^{b\cos\theta} e^{\frac{ikx^2}{2s_0}} dx \quad 20$$

A change of variable of $x = z\cos\theta$ leads to

$$U(P) = -\frac{iA}{\lambda} \frac{e^{ik(r_0+s_0)}}{r_0 s_0} \cos\theta \int e^{\frac{iky^2}{2s_0}} dy \int_a^b e^{\frac{ikz^2}{2s_0}} e^{-\frac{ikz^2 \sin^2\theta}{2s_0}} dz \qquad 21$$

which is the same as equation 18. Thus the approximation of projecting into the plane has the same level of accuracy as the Fresnel approximation itself and may be used with confidence.

At this point the notation is changed from that of Born and Wolf to one that is a little more intuitive for the application at hand. Figure 3 defines the coordinate system, and a list of variables is provided in Table II. F is the distance from mask to focal plane (formerly $s_0$). $\rho$ is the radius on the mask ($\sqrt{x^2+y^2}$), and $\theta$ its azimuthal angle. $s$ is the distance off axis on the focal plane. Then, following the Fresnel approximation for large F

$$E = \frac{E_0 e^{ikF} e^{\frac{iks^2}{2F}}}{i\lambda F} \int_0^\infty e^{\frac{ik\rho^2}{2F}} \rho \int_0^{2\pi} A(\theta,\rho) e^{\frac{ik\rho s \cos\theta}{F}} d\theta d\rho \qquad 22$$

Table II
List of Variables in Fresnel Approximation

| | |
|---|---|
| $\lambda$ | wavelength of light |
| k | $2\pi/\lambda$ |
| $\rho$ | radius of position on starshade |
| $\theta$ | angle of position on starshade |
| E | Electric field amplitude at telescope plane |
| $E_0$ | Electric field amplitude incident on starshade |
| R | residual electric field amplitude in shadow |
| $\varepsilon$ | small dimensionless perturbation factor |
| A | apodization function of starshade |
| F | distance starshade to telescope |
| a | offset radius of hypergaussian |
| $\alpha$ | $a\sqrt{k/F}$ |
| b | 1/e radius of hypergaussian |
| $\beta$ | $b\sqrt{k/F}$ |
| n | order of hypergaussian |
| $\tau$ | $\rho\sqrt{k/F}$ |
| s | distance off optic axis in telescope plane |
| $\sigma$ | $s\sqrt{k/F}$ |
| P | number of petals |

In the case of a circularly symmetric apodization one can first integrate over angle, finding

$$E = \frac{E_0 k e^{ikF} e^{\frac{iks^2}{2F}}}{iF} \int_0^\infty e^{\frac{ik\rho^2}{2F}} A(\rho) J_0\left(\frac{k\rho s}{F}\right) \rho d\rho \qquad 23$$

If A(ρ) is unity to some radius *a*, and zero beyond, and if ikρ²/2F is small, then this integral leads to the familiar Airy disk that describes the point spread function of the typical diffraction-limited telescope.

### *D. On-Axis Analysis*

For mathematical simplicity first confine the analysis of the on-axis (s=0) position. When *s* is much smaller than *F/(kρ)* across the mask, the Bessel function term remains close to unity and equation 23 simplifies to

$$E = \frac{k}{iF} e^{ikF} \int_0^\infty A(\rho) e^{\frac{ik\rho^2}{2F}} \rho d\rho \qquad 24$$

One then seeks a solution that satisfies equation 7, such that

$$\frac{k}{iF} \int_0^\infty A(\rho) e^{\frac{ik\rho^2}{2F}} \rho d\rho = 1 \qquad 25$$

because the phase is unimportant to the depth of the shadow and the term $e^{ikF}$ cancels out.

To investigate an apodization function of the form of equation 2 again use the Fresnel integral as in equation 8

$$E = \frac{k}{iF} \int_0^a e^{\frac{ik\rho^2}{2F}} \rho d\rho + \frac{k}{iF} \int_a^\infty e^{-\frac{(\rho-a)^n}{b^n} + \frac{ik\rho^2}{2F}} \rho d\rho \qquad 26$$

To show this, first perform a change of variable to what turns out to be a set of natural units. Multiplying each distance variable by the same scaling factor gives

$$\alpha = a\sqrt{\frac{k}{F}} \qquad \beta = b\sqrt{\frac{k}{F}} \qquad \tau = \rho\sqrt{\frac{k}{F}} \qquad \sigma = s\sqrt{\frac{k}{F}} \qquad 27$$

so that

$$E = \frac{1}{i}\int_0^{\alpha} e^{\frac{i\tau^2}{2}} \tau d\tau + \frac{1}{i}\int_{\alpha}^{\infty} e^{\frac{i\tau^2}{2} - \left(\frac{\tau-\alpha}{\beta}\right)^n} \tau d\tau \qquad 28$$

and

$$E = 1 - e^{\frac{i\alpha^2}{2}} + \frac{1}{i}\int_{\alpha}^{\infty} e^{\frac{i\tau^2}{2} - \left(\frac{\tau-\alpha}{\beta}\right)^n} \tau d\tau \qquad 29$$

Integration by parts then gives us

$$E = 1 - e^{\frac{i\alpha^2}{2}} - e^{-\left(\frac{\tau-\alpha}{\beta}\right)^n} e^{\frac{i\tau^2}{2}}\Big|_{\alpha}^{\infty} - \frac{n}{\beta}\int_{\alpha}^{\infty} e^{\frac{i\tau^2}{2}} e^{-\left(\frac{\tau-\alpha}{\beta}\right)^n} \left(\frac{\tau-\alpha}{\beta}\right)^{n-1} d\tau \qquad 30$$

or

$$R = \frac{n}{\beta}\int_{\alpha}^{\infty} e^{\frac{i\tau^2}{2}} e^{-\left(\frac{\tau-\alpha}{b}\right)^n} \left(\frac{\tau-\alpha}{\beta}\right)^{n-1} d\tau \qquad 31$$

where E is replaced by R to indicate it is the residual filed inside the shadow.

To evaluate this integral once again integrate by parts:

$$R = e^{\frac{i\tau^2}{2}} e^{-\left(\frac{\tau-\alpha}{\beta}\right)^n} \left(\frac{\tau-\alpha}{\beta}\right)^{n-1} \left(\frac{n}{i\tau\beta}\right)\Big|_{\alpha}^{\infty} + \int_{\alpha}^{\infty} e^{\frac{i\tau^2}{2}} e^{-\left(\frac{\tau-\alpha}{\beta}\right)^n} f(\tau) d\tau \qquad 32$$

where

$$f(\tau) = \frac{n}{\beta^2}\left(\frac{\tau-\alpha}{\beta}\right)^{2n-2}\left(\frac{1}{i\tau}\right) - \frac{n}{i\tau^2\beta}\left(\frac{\tau-\alpha}{\beta}\right)^{n-1} + \frac{n(n-1)}{i\tau\beta^2}\left(\frac{\tau-\alpha}{\beta}\right)^{n-2} \qquad 33$$

The first term of equation 32 is identically zero when evaluated from $\alpha$ to $\infty$, as will be any term that contains both the exponential and a term of positive power in $(\tau-\alpha)/\beta$. Equation 33 has three terms, each of which must be integrated in the second term of equation 32. The first term of equation 33 has a higher power in $(\tau-\alpha)/\beta$ and as such will be a smaller term than the rest of R. The second term is similarly related to R itself, but is smaller by a factor of $n/\tau^2$. Thus, if $\beta^2$ is larger than *n* the third term will dominate. If $\beta^2$ is not larger than *n*, then the transmission rises so quickly near $\tau=\alpha+\beta$ that the shade will start to resemble a disk, and Arago's Spot will re-emerge.

Proceeding to integrate by parts and take the dominant term until a final term that does not evaluate to zero is reached, and the result is

$$R = \frac{n!}{\beta^n} \int_\alpha^\infty e^{\frac{i\tau^2}{2}} e^{-\left(\frac{\tau-\alpha}{\beta}\right)^n} \tau^{1-n} d\tau \qquad 34$$

To approximate the value consider that cosine terms vary rapidly and will integrate to a net of zero at some point in the first half cycle. That cycle will have a length of no more than $1/\alpha$. During this half cycle the second exponential term remains near one and the term in powers of $\tau$ will never exceed $\alpha(1-n)$. So it is expected that

$$R \leq \frac{n!}{\beta^n} \frac{1}{\alpha} \left(\frac{1}{\alpha}\right)^{n-1} = \frac{n!}{\alpha^n \beta^n} \qquad 35$$

which tells the level to which the electric field can be suppressed. The square of R is approximately the contrast ratio to be expected in the deep shadow.

In order to achieve this simplification those terms in the repeated integration by parts that were shown to be small compared to the dominant terms were dropped. Yet in certain parts of parameter space these very same terms can be dominant. For example, as *n* becomes large, the shape of the occulter approaches a circle and the spot of Arago becomes strong again. The validity of this formulation has been checked computationally and found to be reasonable when $\beta^2 > n$. An example of the comparison can be found in Figure 4.

It is clear from inspection of Equation 35 that the greatest suppression of diffraction of an occulter of radius $\alpha+\beta$ (to its inflection point) will occur when $\alpha$ is approximately equal to $\beta$. Also, to achieve high contrast, $\alpha^n$ must be quite large. This is clearly easier to achieve as *n* increases, explaining why the higher order curves give more compact solutions, just a few half zones wide. If *n* gets too high, there are diminishing returns as *n!* rises and $\beta$ approaches unity. Powers as high as *n*=10 or 12 can be practical but n=6 is usually close to providing the widest shadow at a given level of suppression.

Equation 35 also shows that the depth of the central shadow is proportional to $\lambda^{2n}$, which is typically $\lambda^{12}$ for a well-designed starshade. So a practical design will usually be optimized at the longest needed wavelength. Shortward, the performance improves rapidly, while longward the performance very rapidly degrades. This effect is shown numerically in Figure 6. The effect is a property of the offset hypergaussian apodization function that not all other functions exhibit.

### E. Off-Axis Analysis

Consider equation 34, which gives the dominant term of the residual electric field in the center. The diffracted light which reaches the center is mostly coming from the first half cycle of the first term in the integral and is thus coming from a narrow ring just outside $\tau=\alpha$.

Then return to equation 23, but this time include general values of s. The $J_0$ term does not vary significantly across the narrow ring at the edge and may, therefore to excellent approximation, be brought outside the integral, giving us

$$E = \frac{E_0 e^{ikF} e^{\frac{iks^2}{2F}}}{i\lambda F} J_0\left(\frac{kas}{F}\right) \int_0^\infty e^{\frac{ik\rho^2}{2F}} A(\rho)\rho d\rho \qquad 36$$

which leads through the same integration process to

$$R(\sigma) = \frac{n!}{\beta^n} J_0(\alpha\sigma) \int_\alpha^\infty e^{\frac{i\tau^2}{2}} e^{-\left(\frac{\tau-\alpha}{\beta}\right)^n} \tau^{1-n} d\tau \qquad 37$$

What this shows is that, aside from a modulation introduced by the angular integral, the residual electric field remains the same. In other words, the field at any point off axis is dominated by the diffraction at the nearest edge. Given how quickly the diffraction rises off axis, this is not unexpected.

Finally, consider that the integral is from $\alpha$ to $\infty$. There is no contribution from closer to the center than $\alpha$. So, until one passes the center and starts approaching the other side, $\alpha$ is simply the measure of how far underneath the opaque section the point lies. Consequently, one can rewrite equation 35 as

$$R \leq \frac{n!}{(\alpha-\sigma)^n \beta^n} \qquad 38$$

Or, redefining ($\alpha$-$\sigma$) as $\gamma$ (the distance inward from the effective edge at $\alpha$) one finds

$$R \leq \frac{n!}{\gamma^n \beta^n} \qquad 39$$

where $\gamma$ is at least somewhat greater than unity.

So, the shadow can be understood (approximately) as starting with intensity of $\beta^{-2n}$ just inside the opaque circle and then falling as $\gamma^{-2n}$ down to the center.

## III. Two-Dimensional Computer Modeling

The most obvious approach to the problem of computer computation is simply to evaluate the Fresnel integral (equation 26) directly at each point in the shadow. Unfortunately, the number of points to be evaluated before the accuracy of the net integral reaches the required suppression level of $R^2$ is on the order of $R^{-2}$. So a single point in the shadow plane can require a trillion sine calculations at quadruple precision. Because the direct approach becomes impractically slow, alternative, faster approaches are required.

At least three such codes have been developed by members of the New Worlds team: the edge integral approach discussed here, a code that performs a Fourier propagation of the Fresnel diffraction (Glassman et al, 2009) and a Hankel Transform (Vanderbei, Cady and Kasdin, 2007).

A physically oriented code is desirable, particularly for tolerance simulations where a small deviation can be added or subtracted on its own, without being convolved with the rest of the system. Such an approach makes direct use of the fundamentally binary nature of the starshades. All parts of the starshade must be either fully opaque or fully transmitting. Errors are thus related to errors in the projected shape as defined by the outline of the occulter.

A solution that would operate in a manner similar to a Green's Theorem, in which a surface integral can be converted to a line integral around the edge would be ideal. Dubra and Ferrari (1999) published a paper entitled "Diffracted field by an arbitrary aperture" in which they integrated the Kirchoff formulation of diffraction theory by means of a Green's function approach and converted the two-dimensional integral to a one-dimensional parametric integral. Their approach is adopted here, but only in the simpler case of a plane wavefront.

In the case of a binary optic, the apodization is everywhere unity across the aperture, so that equation 22 becomes

$$E = \frac{E_0 e^{ikF} e^{\frac{iks^2}{2F}}}{i\lambda F} \iint_S e^{\frac{ik\rho s \cos\theta}{F}} e^{\frac{ik\rho^2}{2F}} \rho \, d\theta \, d\rho \qquad 40$$

where $S$ represents the surface of the aperture. But S is a completely general surface, and, specifically, there is no requirement that the surface be centered or symmetrical about the origin. So, if the source is at infinity, an off-axis point is calculated by moving the aperture off center. That allows $s$ to be set to 0 for any point in the focal plane, by shifting the aperture of integration.

So, setting $E_0$ to unity

$$E = \frac{e^{ikF}}{i\lambda F} \iint_S e^{\frac{ik\rho^2}{2F}} d\theta \rho \, d\rho \qquad 41$$

and, integration over ρ in closed form yields

$$E = \frac{1}{2\pi} \int_0^{2\pi} e^{\frac{ik\rho^2}{2F}} d\theta \qquad 42$$

evaluated from the inner radius $\rho_i$ to the outer radius $\rho_o$ at each value of θ.

In the case where the area does not include the origin, and is simple, in that any radial, non-osculating line cuts the surface twice, the result is

$$E = \frac{1}{2\pi} \int_0^{2\pi} e^{\frac{ik\rho_o^2}{2F}} d\theta - \frac{1}{2\pi} \int_0^{2\pi} e^{\frac{ik\rho_i^2}{2F}} d\theta \qquad 43$$

In the case where the area is simple, and the origin is inside, then each radial line cuts the perimeter once at $\rho_o$ and $\rho_i$ is everywhere 0, so

$$E = \frac{1}{2\pi} \int_0^{2\pi} e^{\frac{ik\rho_o^2}{2F}} d\theta - 1 \qquad 44$$

In equation 43 the first term is the line integral along the far edge of the area, while the second term is the return on the near side. Thus the integral can be turned into a line integral around the edge of the shape. So, in the case of a simple, convex shape that excludes the origin ($\rho=0$) within, the equation becomes

$$E = \frac{1}{2\pi} \int_s e^{\frac{ik\rho^2}{2F}} \frac{\hat{\rho} \cdot \vec{ds}}{\rho} \qquad 45$$

where $\hat{\rho}$ is the unit vector in the radial direction and $\vec{ds}$ is in the direction of the normal to the edge element and has size equal to length of the edge element. So one merely breaks the edge into small elements and sums the phase factor around the edge.

In the case where the shape is simple, but includes the origin inside

$$E = \frac{1}{2\pi} \int_s e^{\frac{ik\rho^2}{2F}} \frac{\hat{\rho} \cdot \vec{ds}}{\rho} - 1 \qquad 46$$

From an algorithmic point of view, a simple prescription for the electric field at the origin emerges. Create a set of points that outline the starshade. At each point calculate the distance between the adjacent points and create the vector *ds*, which is the vector normal to the surface at that point, with a value equal to the length of the edge element. For each element create the dot product of the normal and the unit vector from the center. Divide by distance from the center and multiply by the Fresnel phase term. Sum this all the way around the edge, and the result will be the desired value in the center. To find a point off axis, shift the shape terms and recalculate.

It should be noted that this works well for non-simple forms as well. A complex shape may be broken into simple shapes and each shape integrated separately. The borders between the simple shapes are integrated in one direction for one shape and in the other direction for the adjacent shape, so the net along the border is zero. In practice this means that one can follow the algorithm described in the preceding paragraph around the edge of any, arbitrary shape. Holes may be calculated inside a mask by integrating the edge in the opposite direction. Of course, one must still calculate whether or not the origin falls inside or outside the shape. If it is found to be inside, then one must subtract the one.

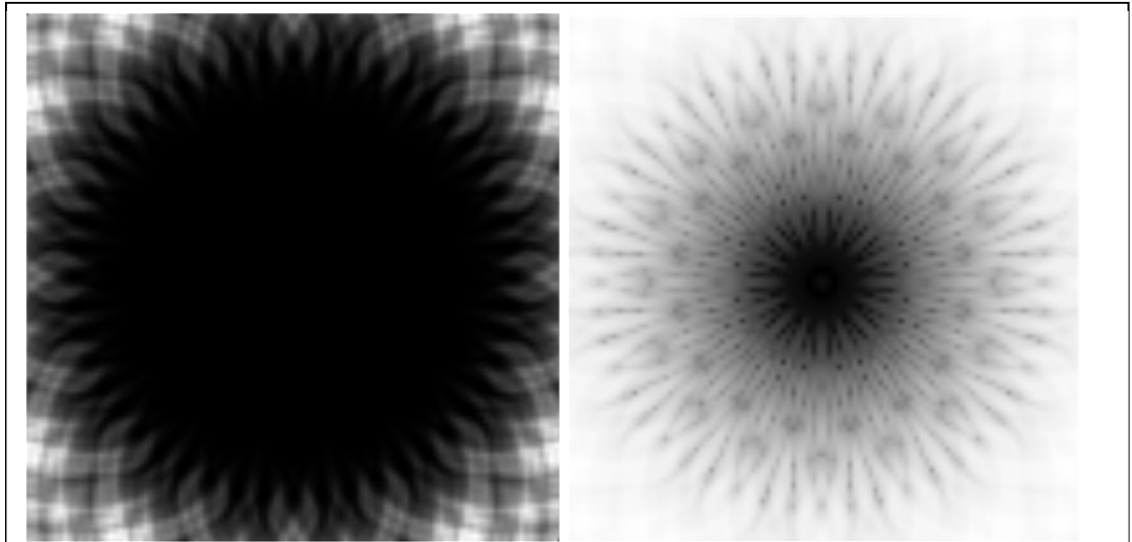

Figure 4: The suppression caused by a starshade (a=b=12.5m, n=6, F=80,000km, λ=0.5μm) is shown in the shadow plane. An array of points 128square was calculated across a 50x50m square in the plane of the telescope. To the left is the intensity of the residual shadow on a linear scale, showing complicated diffraction patterns near the edge and a fast fall-off to the center. To the right is the same shadow diagram plotted on logarithmic scale, showing more complex structure and a very deep shadow toward the center.

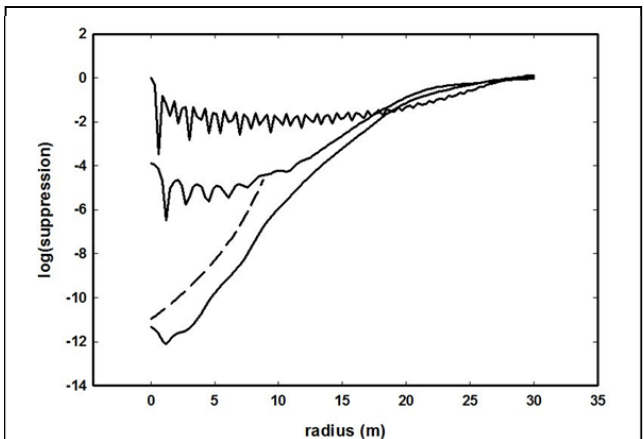

Figure 5: The suppression caused by starshades is shown as a function of shadow radius. All four curves feature a starshade of radius 25m at 80,000km operating at a wavelength of 0.5μ. The top line is for a simple disk and shows the spot of Arago at the center where the suppression vanishes. The next curve down is for a simple Gaussian shape with no offset and 16 petals. The bottom curve is for an offset hypergaussian with a=b=12.5m and n=6, showing suppression down to well below $10^{-10}$. The dashed line is the approximation of equation 38, which shows that it tends to err on the conservative side.

Such a code was built and it works very effectively, and very quickly. It typically takes 0.1 seconds on today's laptops to calculate a single point in the shadow. About 40,000 points are needed around the edge of a starshade to gain sufficient accuracy to predict the residual field to the $10^{-12}$ level. At the start of the algorithm the starshade is defined through four vectors. These are the x and y values of the points around the edge and the x and y values of the normal vectors.

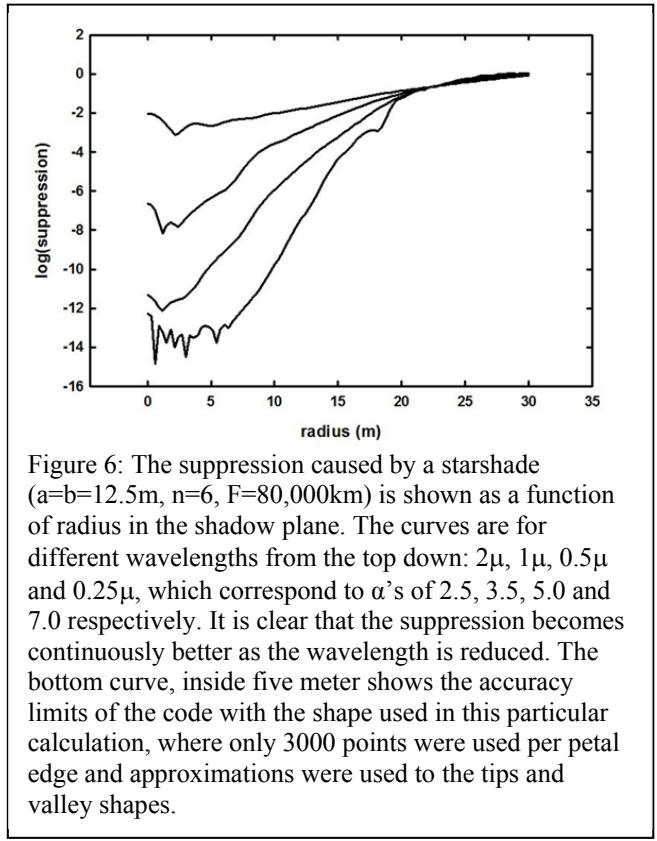

Figure 6: The suppression caused by a starshade (a=b=12.5m, n=6, F=80,000km) is shown as a function of radius in the shadow plane. The curves are for different wavelengths from the top down: 2μ, 1μ, 0.5μ and 0.25μ, which correspond to α's of 2.5, 3.5, 5.0 and 7.0 respectively. It is clear that the suppression becomes continuously better as the wavelength is reduced. The bottom curve, inside five meter shows the accuracy limits of the code with the shape used in this particular calculation, where only 3000 points were used per petal edge and approximations were used to the tips and valley shapes.

The trickiest part of the algorithm is finding a way to accurately check whether or not the origin is inside the shape. This is difficult near the edges where there is a mathematical discontinuity, and an incorrect value of inside/outside can lead to a false value of E near unity, when the true value may be very different. It is even more difficult near the corners and tips of the shade. The vectors must be built with care there to ensure that small, round-off errors do not create incorrect values for the inside/outside determination.

In Figure 4 we show a map created with this code by calculating the intensity in a 128x128 array of shadow plane points for an offset hypergaussian starshade with a=b=12.5m, n=6, at F=80,000km and λ=0.5μ. With sixteen petals, this starshade creates complicated, two dimensional patterns but also creates the deep central shadow desired.

In Figure 5 we plot the average radial intensity of the same starshade, and compare it to the performance of a simple disk and a simple Gaussian. We also show the prediction of Equation 38 and see that the simple formulation tends to err on the conservative side.

Figure 6 is the same hypergaussian starshade evaluated at four different wavelengths, showing that the performance continues to improve as wavelength decreases. We can also see some inaccuracy from numerical integration down near the $10^{-14}$ level.

The code is versatile because it mimics reality rather closely. A small deviation from the nominal value of the edge in reality is reflected directly in the sum of the residual electric field. The code sums the local behaviors to create a single global value at a point. This makes the code ideal for modeling tolerances and other real effects. In Figure 4 the code is used to calculate the depth of the shadow as a function of radius for a 16 petal starshade and compare it to the circularly symmetric approximation. The results have been carefully cross-checked with another code that has been reported upon elsewhere (Glassman et al, 2009).

## IV. Tolerancing

So far, the starshade concept has been treated as a mathematical construct, without regard to its practical application. But if it is ever to be built, the tolerances for fabrication must be investigated. Any device in which the tolerances are impractically tight would not be achievable and thus would be of little value. It is the purpose of engineering studies to determine what is actually achievable and at what cost. Many such studies (Shipley et al, 2007; Lyon et al, 2007; Arenberg et al, 2008, Dumont et al, 2009; Kasdin et al, 2009; Shaklan et al, 2010) have now been performed and the community has a rough idea of where the boundaries of practicality and affordability lie. That there is a general sense that the tolerances can be met in affordable programs is actually the greatest strength of starshades.

The tolerance discussion is started with an inspection of equation 42 . While this equation was generated while searching for a method of numerical simulation, it is very useful for discussing tolerances. First convert the equation to dimensionless, natural units using the definitions of equation 27, so that:

$$E = \frac{1}{2\pi} \int_0^{2\pi} e^{\frac{i\tau^2}{2}} d\theta \qquad 47$$

where it is understood that τ is given as a function of θ.

Through change of variable and use of the chain rule this equation reads

$$E = \frac{1}{2\pi} \int_0^\infty e^{\frac{i\tau^2}{2}} \frac{d\theta}{d\tau} d\tau \qquad \qquad 48$$

So that now the outline of the shade is defined by radius as a function of angle. $\tau(\theta)$ does not need to be single-valued. The integral is simply executed over all values of $\tau$ at any $\theta$.

The presence of the $d\theta/d\tau$ term gives insight into the tolerancing of a binary optic. Large leaps and discontinuities in $\tau$ can be tolerated as long as $d\theta/d\tau$ remains zero. But a discontinuous change in $\tau$ means a linear edge that points directly at the shadow point under evaluation.

If that edge is misaligned with the point of evaluation (e.g. an off-axis point) then large amounts of diffraction can rapidly develop. For example, if the edge covers one half-zone, then the change in electric field is

$$\delta E \approx \frac{1}{2\pi} e^{\frac{i\tau^2}{2}} \delta\theta \approx \frac{\delta\theta}{2\pi} \qquad \qquad 49$$

where $\delta\theta$ is the projected angle of the edge as viewed from the center. If $10^{-10}$ contrast is desired, then $\delta E$ must be held to $10^{-5}$ and $\delta\theta$ must then be below about $10^{-4}$. For a 50m diameter shade, the resultant shadow would be only 5mm in diameter. This effect is clearly seen as a reduction in the diameter of the deepest part of the shadow as a function of petal number in Figure 8.

It should be noted that in the starshade designs, the diameter of the shadow is much larger than this. The perimeter of the starshade is closest to radial at the tips and in the valleys near the base. At each of these points there is a nearby matching edge at the same angle and in the opposite direction. To first order they cancel as $\delta\theta$ grows. To higher order, $\delta\theta$ is not linear and the Fresnel phase is not exactly the same on either side and can play a small role in the off-axis response.

Letting $\tau(\theta)$ be perturbed by a function $\varepsilon(\theta)$:

$$E + \Delta = \frac{1}{2\pi} \int_0^{2\pi} e^{\frac{i(\tau+\varepsilon)^2}{2}} d\theta \qquad \qquad 50$$

where Δ is now the change in the electric field in the shadow. Expanding and dropping higher terms then gives:

$$\Delta = \frac{i}{2\pi} \int_0^{2\pi} e^{\frac{i\tau^2}{2}} \tau \varepsilon \, d\theta \qquad 51$$

as a general measure of the effect of an error. It should also be noted that this can be changed from an error function ε in the radial direction to an error function δ in the azimuthal direction yielding

$$\Delta = \frac{i}{2\pi} \int_0^\infty e^{\frac{i\tau^2}{2}} \tau \delta(\tau) \, d\tau \qquad 52$$

Inspection of equation 52 shows that an error of the form

$$\delta(\tau) = \delta_0 e^{-\frac{i\tau^2}{2}} \qquad 53$$

is about as bad as possible, creating an effect of size $\Delta \sim \tau \delta_0$, where τ is roughly the length of the error along the edge. Similarly, an error that is localized within one Fresnel zone will cause an error $\Delta \sim \tau \delta_0$, where $\tau \delta_0$ is the area of deformity in outline.

### A. Petal Number

It is remarkable (and not fully intuitive), but a circularly symmetric apodization function may be well approximated by petals (Figure 7), allowing the occulter to be binary (Marchal, 1985). While strictly speaking the number of petals is a design choice, not a tolerance, analysis of petal number follows in the form of a simple tolerance analysis. In part II of this paper a circularly symmetric formulation for the apodization function was used, which would have required a partially transmitting aperture. In practice, scattering from the transmitting material would keep such designs from being easily built. A binary optic with a finite number of petals is required. It has been established through raytracing (as discussed in section III) that, for the design range in use, 16 petals provides an approximation to circularly symmetry with no major loss of performance (Glassman et al, 2009). Twelve petals can be used at the expense of some loss of deep-shadow diameter. Below that, the size of the shadow shrinks rapidly with petal number.

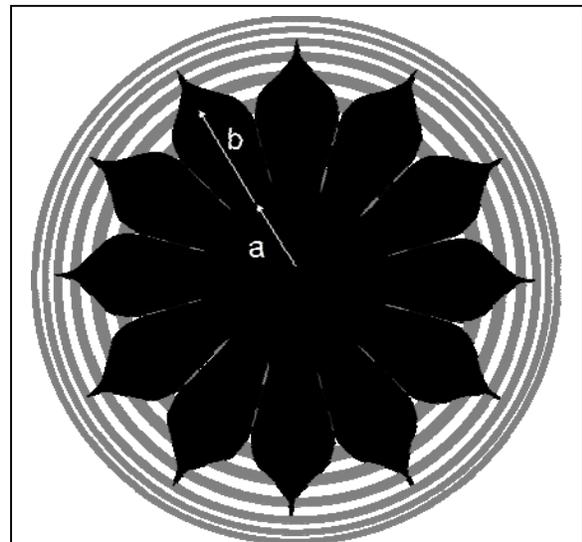

Figure 7: A twelve petal version of the starshade is shown schematically with Fresnel zones in the background.

The reason for this can be understood from examination of equation 47. Moving off axis by a distance δτ toward a single petal results in a very strong increase in diffraction as discussed earlier, even in the case of an infinitely narrow petal. However, moving a distance δτ perpendicular to a petal causes a much smaller effect, creating an ε given by

$$\varepsilon = \tan^{-1}\left(\frac{\delta\tau}{\tau}\right) \qquad 54$$

which, when δτ/τ is small, gives us

$$\Delta = \frac{i}{2\pi}\int_0^{2\pi} e^{\frac{i\tau^2}{2}} \tau\left(\frac{\delta\tau}{\tau}\right) d\theta = (\delta\tau) E \qquad 55$$

So E becomes (1+δτ)E, which is a small effect.

However, when the small angle approximation of the arctan in equation 54 starts to break, at values that become a significant fraction of π/2, then the errors start to grow rapidly. At π/6 the approximation is quite good, indicating twelve petals is reasonable. Calculations were made with the code discussed in section III and are shown in Figure 8. They show that the central spot and the areas near the edge of the shadow are not significantly impacted by petal number, but below twelve to sixteen petals the size of the central dark shadow decreases rapidly.

An important point about petal-shaped shades can be easily shown from these equations: Each petal operates independently. In particular, the diffraction from one side of the shade is not used to cancel the light from the other side. Similarly, there is no need for uniformity of design from one petal to the next.

Consider rewriting equation 47 to reflect its petal nature. If the shade has P identical petals then

$$E = \frac{1}{2\pi P} \sum_{i=1}^{P} \int_{\frac{2\pi(i-1)}{P}}^{\frac{2\pi i}{P}} e^{\frac{i\tau^2}{2}} d\theta \qquad 56$$

By symmetry each petal is the same so each element of the sum is identical and thus $E_i = E/P$, where $E_i$ is the contribution for the $i^{th}$ petal. Each petal individually sums to zero.

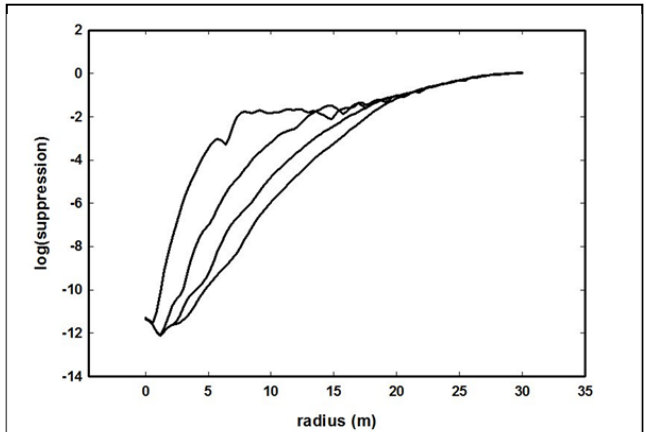

Figure 8: The suppression caused by a starshade (a=b=12.5m, n=6, F=80,000km, λ=0.5μm) is shown as a function of radius in the shadow plane. The curves are calculated for different numbers of petals. From the top down there are 4, 8, 12 and 16 petals respectively. One sees that the edges and the center of the shadow are not affected by petal number, but the size of the central hole is significantly compromised below about 12 petals.

Thus the parameters of each petal may vary. In particular, its width and length may vary as long as each $E_i$ still remains acceptably small. Asymmetries, however, can have some effects on tolerances and field of view, so breaking symmetry must be done with care.

**B.**  **Alignment**

*Lateral Position:* This is the position of the detector perpendicular to the line that extends from the source through the center of the starshade. If the telescope drifts too far laterally, it will start to leave the shadow. This distance is set by the size of the shadow. The depth of the shadow increases as one approaches the center, and the telescope must be smaller than the diameter of the region with sufficient contrast. This region becomes larger as the shade becomes larger and more distant. Thus, an optimized starshade would fit the shadow size to the telescope size. So, a margin of 20% on the starshade size appears reasonable. Thus simply choose $\pm 0.1a$ as the constraint on lateral position.

*Depth of Focus:* This is the position of the detector along the line from the star through the center of the starshade. There is no focal plane for the telescope in the shadow as it is deep along its entire length. However, as the telescope moves farther from the starshade along the shadow, two things happen - the inner working angle drops and the amount of diffracted light rises. So the depth of focus is set by a trade between these two effects. Equation 35 shows that the residual diffraction shadow scales as $F^{2n}$. Since n is typically 6, the residual diffraction will rise as the twelfth power of the distance. Even a one percent increase in distance could lead to a detectable (12%) increase in diffraction. On the other hand, a one percent change in inner working angle is usually not serious. So the position of the telescope should be known to 1% in the beam (800km in our standard case) and this position tolerance does not present a serious difficulty.

*Rotational*: Because of the circular symmetry built into the design, there is no constraint on $\theta_z$, the rotation angle about the line of sight. Sometimes it might be better to actually spin the starshade about this axis to smooth out residual diffraction effects.

*Pitch and Yaw:* Because of the rotational symmetry the constraint on errors in alignment about the pitch axis, $\theta_x$ and yaw axis, $\theta_y$, may be combined into a single pointing error. It turns out that the design is highly forgiving of such errors, but the proof takes some calculation.

Assume that the shade is out of alignment with the axis of symmetry by an angle $\varphi$ about the y-axis, such that the shade appears foreshortened in the x direction by a factor of $\cos\varphi$, which is approximated by $1-\varepsilon$. The net optical path difference is small, about $(a+b)\theta\varphi^2/2$ for small $\theta$ and $\varphi$. As long as $\varphi$ is $<<1$ the net path delay is a small fraction of a wavelength and may be ignored.

If this is not the case, then start by rewriting equation 24 for the on-axis (s=0) case in Cartesian coordinates with the integration now taking place over the projected area which is foreshortened in one dimension

$$E = \frac{k}{2\pi i F} e^{ikF} \left[ \begin{array}{l} \int e^{\frac{ikx^2}{2F}} \int e^{\frac{iky^2}{2F}} dxdy + \\ \\ \int e^{\frac{ikx^2}{2F}} \int e^{\frac{iky^2}{2F}} e^{-\left(\frac{\sqrt{x^2+y^2}-a}{b}\right)^n} dxdy \end{array} \right] \qquad 57$$

By a change of coordinate to z=x/(1-ε)

$$E = \frac{k}{2\pi i F} e^{ikF} \left[ \begin{array}{l} (1-\varepsilon) \int e^{\frac{iky^2}{2F}} \int e^{\frac{ikz^2(1-\varepsilon)^2}{2F}} dydz + \\ \\ (1-\varepsilon) \int e^{\frac{iky^2}{2F}} \int e^{\frac{ikz^2(1-\varepsilon)^2}{2F}} e^{-\left(\frac{\sqrt{y^2+z^2(1-\varepsilon)^2}-a}{b}\right)^n} dydz \end{array} \right] \qquad 58$$

where the integration is now over a circularly symmetric shape as before. Converting to polar coordinates

$$E = \frac{k}{2\pi i F} e^{ikF} \left[ \begin{array}{l} (1-\varepsilon) \int_0^{2\pi} \int_0^a e^{\frac{ik\rho^2}{2F}} e^{-\frac{ik\rho^2 \cos^2\theta(2\varepsilon-\varepsilon^2)}{2F}} \rho d\rho d\theta + \\ \\ (1-\varepsilon) \int_0^{2\pi} \int_a^\infty e^{\frac{ik\rho^2}{2F}} e^{-\frac{ik\rho^2 \cos^2\theta(2\varepsilon-\varepsilon^2)}{2F}} e^{-\left(\frac{\sqrt{\rho^2-\rho^2\cos^2\theta(2\varepsilon-\varepsilon^2)}-a}{b}\right)^n} \rho d\rho d\theta \end{array} \right] \qquad 59$$

Expanding and ignoring terms in $\varepsilon^2$ and higher, then differencing from the unperturbed integral yields an expression for the remainder caused by the misalignment:

$$R = \frac{k}{2\pi i F} \int_0^{2\pi} \int_0^a e^{\frac{ik\rho^2}{2F}} \left[ 1 - (1-\varepsilon)e^{-\frac{ik\varepsilon\rho^2 \cos^2\theta}{2F}} \right] \rho d\rho d\theta +$$

$$\frac{k}{2\pi i F} \int_0^{2\pi} \int_a^\infty e^{\frac{ik\rho^2}{2F}} e^{-\left(\frac{\rho-a}{b}\right)^n} \left[ 1 - (1-\varepsilon)e^{-\frac{ik\varepsilon\rho^2 \cos^2 \vartheta}{2F}} e^{\left(\frac{\rho-a}{b}\right)^n \left(1 - \left(1 - \frac{\rho\varepsilon\cos^2\theta}{\rho-a}\right)^n\right)} \right] \rho d\rho d\theta \qquad 60$$

Approximation of the exponentials in the brackets and dropping higher order terms reduces this to:

$$R = \frac{\varepsilon k}{2\pi i F} \int_0^{2\pi}\int_0^a e^{\frac{ik\rho^2}{2F}} \left[1 + \frac{ik\rho^2 \cos^2\theta}{F}\right] \rho \, d\rho \, d\theta +$$

$$\frac{\varepsilon k}{2\pi i F} \int_0^{2\pi}\int_a^\infty e^{\frac{ik\rho^2}{2F}} e^{-\left(\frac{\rho-a}{b}\right)^n} \left[1 + \frac{ik\rho^2 \cos^2\theta}{F} - \left(\frac{\rho-a}{b}\right)^n \left(\frac{n\rho \cos^2\theta}{\rho-a}\right)\right] \rho \, d\rho \, d\theta \qquad 61$$

The terms in higher order of ρ are smaller as before, leaving an expression for the remainder. To first order, the remaining electrics field $R_\theta$ is given by

$$R_\theta = (1 - \cos\theta) R \approx \frac{\theta^2}{2} R \qquad 62$$

where R is the residual electric field in the original untilted case. So misalignments of axis will not be severe and many degrees of misalignment can be tolerated.

### C. Tips and Valleys

*Truncation of Petals:* Mathematically, the apodization carries out to infinity. In the case of a binary mask, this means that petals extend to infinity, something which clearly cannot be done in practice. At what radius is it safe to truncate the petal? One can write the remainder of the electric field created by truncating at a radius T.

$$R = \int_T^\infty e^{\frac{i\tau^2}{2}} e^{-\left(\frac{\tau-\alpha}{\beta}\right)^n} \tau \, d\tau \qquad 63$$

which is definitely less than

$$R = \frac{1}{P} e^{-\left(\frac{T-\alpha}{\beta}\right)^n} \qquad 64$$

per petal. The remainder due to truncation can be safely ignored in a typical case when the thickness of each petal has fallen below about 0.1mm. Thus the petals must be sharp at their tips, but do not have to be controlled at a microscopic level.

### D. Distortions

*Area Change:* Consider the case where the shape changes in a discontinuous manner. Since there are many possible classes of such error, they can only be addressed as a generality. Consider a petal that is missing a chunk along one edge. The missing part can be contained within one half zone or spread over several. To the extent that the missing area is monotonic across the zones, the net effect is less than the largest area within one half zone. So, the size of the missing area must be less than $10^{-5}$ of the starshade area, but can be substantially larger if spread over several zones.

### E. Shape

*Flatness:* A starshade is not a mirror or a lens and does not alter the phase of a wavefront as it passes by. As such, the flatness requirements are very forgiving. The tolerances are set by the projected shape of the starshade onto the sky. Inside the edge of the frame, the flatness has no effect whatsoever.

Consider the case of an error in which parts of the frame (outlining the sky) move toward or away from the telescope in such a way that the projected shape remains unchanged. Then the field in the shadow may be written as a modification of equation 42:

$$E = \frac{1}{2\pi} \int_0^{2\pi} e^{\frac{ik\rho^2}{2F} + \frac{ik\delta(\theta)\alpha^2(\theta)}{2}} d\theta \qquad 65$$

where $\delta(\theta)$ is the deviation of the shade edge in the z direction as a function of azimuthal angle and $\alpha(\theta)$ is the angular radius of the shade as viewed from the telescope as a function of azimuthal angle.

Assuming that $k\delta\alpha^2$ is much less than unity, the change to E will be given by

$$\delta E = \frac{1}{2\pi} \frac{k\delta_0 \alpha_0^2}{2} \int_0^{2\pi} e^{\frac{ik\rho^2}{2F}} \Phi(\theta) d\theta \qquad 66$$

where $\delta_0 \alpha_0^2$ is the maximum amplitude of the phase delay and $\Phi(\theta)$ is the phasing of the errors around the circumference. Then, noting that the integral cannot exceed $2\pi$ in the worst case, we have

$$\delta E < \frac{k\delta_0 \alpha_0^2}{2} \qquad 67$$

creating a tolerance of

$$\delta_0 < \frac{2R}{k\alpha_0^2} \qquad 68$$

Which means $\delta_0$<2.5 meters in the tightest case. The warp would have to reach ±2.5m excursions on a 1m radial distance to cause detectable degradation. It would take applications in which suppression below $10^{-16}$ is required to make warping a concern.

*Azimuthal Errors in Petal Shape:* When the apodization function was approximated with the petals to make the function binary, the distribution of the electric field was significantly perturbed in the azimuthal direction. The total, when integrated over the circle at any given value of ρ, remained unchanged. Thus, within the azimuthal sector of width 2π/N radians at any fixed radius ρ, the obscuration may be freely moved. Essentially, the starshade is insensitive to shear in the azimuthal direction. Simply keep the shear from slipping into the region of the adjacent petals.

*Radial Errors in Petal Shape:* If the petal is stretched or compressed such that the smoothness of the fall of the apodization is maintained, then there is little impact on the performance. This is reflected in the insensitivity to alignment, wherein the petals in some directions are changed in projected length, but there is no noticeable impact on performance. Similarly the petal analysis shows that each petal independently creates its own deep shadow zone. Hence, radial scaling of modest amounts does not hurt the performance.

### F. Holes

*Opacity:* The shade must be opaque to the needed level. If the star is to be suppressed to better than a ratio S, then the shade must transmit less than 1/S of the incident radiation

*Pinholes:* The presence of pinholes can simulate a level of transparency. By the Fresnel integral the area of the pinholes must represent 1/S of the area of the starshade if uniformly distributed. If contained in one zone, they must add up to less than $1/\sqrt{S}$ of the area of that zone. This tolerance is typically achieved in engineering designs by triple layering the opaque sheet. See, for example, Cash (2009).

*Large Holes:* A single large hole can be restricted to a single zone. Since a zone has an area

$$A_z = \pi \lambda F \qquad 69$$

the hole must have an area less than

$$A_{Hole} < \frac{\pi \lambda F}{\sqrt{S}} \qquad 70$$

which, for typical cases, translates to a hole area as large as a square centimeter, well within a practical range.

### G. Target Constraints

While not strictly a tolerance on the design of the starshade, the properties of the target system can significantly affect the design and operation of a starshade system.

*Stellar Diameter:* The stars we wish to suppress have significant angular extent across the sky. Alpha Centauri's disk is 7 milliarcseconds (mas) in diameter, and our typical target near 10pc will subtend about 1mas. The light from a stellar disk is incoherent, meaning that the shadow will be the convolution of the disk function with the intensity shape of the shadow from a point source. Since the intensity rises so very steeply near the edge it is the rim of the stellar disk that dominates the shadow degradation. A star of diameter θ will cause a diameter loss of Fθ at the telescope. One milliarcsecond at 80,000km creates a 40cm loss in shadow diameter, which should not be forgotten when designing the shade. Essentially, the shade must be made 40cm larger in diameter.

*Seeing*: When light passes through non-uniform, transparent media, phase delays can be introduced as a function of position. When a star is viewed through the atmosphere, these time-variable phase delays cause the image to move around, an effect referred to as "seeing". The phase delays can even split the apparent image of a point into multiple points. Since the phase delays are a coherent effect, the electric field in the telescope plane will be the convolution of the point response electric field with the amplitude of the incident electric field as a function of position on the sky. Because the incident light is coherent, the convolution will include phase effects, unlike the convolution for a stellar disk. But, the electric field is also very steep near the edge, rising typically as the sixth power of radius. So phase effects are quickly overwhelmed by the outlier (in radius) contributions. It is beyond the scope of this paper to discuss the complicated response that is likely to ensue, but the net global result will be similar to the incoherent case. The shadow will be convolved with the seeing disk on the sky. As long as the seeing disk remains within the central suppression zone, the starshade will operate properly. Again, the size should be adjusted in advance to allow for the expected seeing. But a remarkable conclusion is reached: external occulters will work with the atmosphere albeit with an inner working angle several times larger than the seeing.

*Binaries*: Many stars, including our closest neighbor Alpha Centauri, are in binary systems. If the two stars are very close, such that both components lie in the central suppression zone, then observation may proceed as normal. For a widely spaced binary like Alpha Cen, which has zero and first magnitude components separated by about ten arcseconds, suppression of just one component is insufficient. Two independent starshades are required. If the separation of the components is comparable to the inner working angle, such that two occulters are required, but their projected shapes overlap, then the resultant diffraction would be serious and could destroy the suppression. A larger or non-circular shade will be required. Of course, if the nearby source is vastly fainter, like a brown dwarf or exozodiacal light, then it may not pose a problem, depending on the details of the telescope performance.

## V. Apodization Ripples

In 2007, Vanderbei, Cady, and Kasdin (hereafter VCK) published the results of a generalized search for the optimal starshade apodization function. Working with circular symmetry only, they found solutions that have proven to translate well to the petal approximation. Their solutions allow for shrinking the starshade radius by about 25% relative to a hypergaussian design. But the decrease in size is not without cost. Herein is a simple analytic discussion of these somewhat smaller starshades.

Inspection of the plot of the VCK apodization function shows it to be highly similar to an offset hypergaussian. It begins with an opaque center and then falls exponentially to a short tail. Only very close inspection reveals the differences. The biggest difference is a series of ripples on top of the base function. There are some ripples of wavelength comparable to the width of a Fresnel zone that have amplitudes on the order of 1%. There are also some shorter wavelength ripples of magnitude near 0.1%. The other noticeable difference is that the ripples extend closer to the center than in a comparable hypergaussian. In a typical hypergaussian design *a=b*, and no light inside radius *a* is passed.

Consider a hypergaussian that is substantially similar to the rippled function of VCK but fully envelops the bumps. Such a function would give good performance on-axis, but would have a smaller shadow than the VCK case. A large telescope would encounter problems collecting too much diffracted starlight at the edge of the mirror. The ripples can then be thought of as extra apertures opened strategically along the radius to suppress the light around the edge of the shadow. This must be done in such a way that the center of the shadow is not degraded beyond specification. It must also be done in such a way that the broadband response is not lost.

To understand the function of these "apertures", imagine starting with the proximate hypergaussian. In the plane of the telescope mirror, the residual, diffracted electric field may be mapped in strength and phase as a function of radius at any wavelength. The strength increases at a very high rate with radius, and is always the worst at the longest wavelength. Thus, the shadow size improvement starts with the longest wavelength, just outside the radius where the diffracted intensity reaches allowed maximum. Remember that its signal comes almost exclusively from the starshade at a radius of *a*.

To suppress the electric field in the shadow-plane annulus, coherent radiation 180 degrees out of phase must be added. The only source of such radiation is to open an extra aperture in the shade one Fresnel half zone away as viewed from the point in the shadow plane. These points on the shade occur where

$$\frac{(\tau-\sigma)^2}{2} - \frac{(\alpha-\sigma)^2}{2} = \pi + 2\pi n \qquad 71$$

or

$$\tau = \sigma \pm \sqrt{2\pi(1+2n)+(\alpha-\sigma)^2} \qquad 72$$

In a typical application $\alpha \sim 3$ and $\sigma \sim 1$, so the apertures need to be at $\tau = 4$ and $\tau = -2$. The positive solution is located on the sloping edge of the petal, while the other aperture is inside the opaque disk across the center of the starshade. This explains the need to open an aperture inside $\alpha$.

The rest of the apertures are then added to undo the collateral damage from the first aperture. That first one created a Bessel function in the electric field that offset the residual hypergaussian field at $\alpha$. But it also creates a substantial new component of diffracted light near the middle. The additional apertures create additional electric field components designed to offset the new field in the center, but have minimal effect at $\sigma$.

Note that the positioning of these apertures depends on the square root of wavelength and it is thus not surprising that the solution works over a fairly broad band shortward of the design point, but fails eventually. It appears the function of the short wavelength ripples is to extend suppression further to the blue without significantly impacting the red end. The overall bandpass achieved through this means covers more than an octave of spectrum, which is satisfactory for many applications.

However, the use of these discrete features changes the tolerances and fabrication significantly. First, consider that the smallest-radius perturbation on a petal is designed to create a diffractive wave that crosses the axis of the starshade to improve the performance in the shadow of the petal on the other side. One of the highly desirable features of a hypergaussian is that each petal operates independently. The shape and positioning of the petal on one side, does not affect the petal on the other side. Loss of this feature makes fabrication significantly more difficult.

These errors can come about in two ways. They can be the result of a shape error or they can result from positioning errors. Consider that each of these apertures is being convolved with the Fresnel zones. A major ripple (1% of apodization) can move out of position no more than 0.1% of a Fresnel half zone (circa 1mm) relative to the other ripples if $10^{-10}$ suppression is to be maintained. On the other hand, hypergaussians have a smooth shape. Each Fresnel half zone cancels against the next and thus positioning of the shape is more forgiving.

Overall the rippled geometry offers features of interest relative to the hypergaussian. In particular, it allows the diameter of the starshade to be reduced by about 25% without loss of shadow size. Consequently, the inner working angle at which planets are observable can be supported with the starshade 25% closer. A mission may be designed with a savings on both launch mass and maneuvering fuel.

On the other hand, the ripples restrict the bandpass, allowing unacceptable diffraction in the ultraviolet. They also make the fabrication and stability tolerances much more difficult to achieve.

It should be noted that adjustable apertures might be practical. One could literally open or close apertures as needed in flight to correct minor shape errors. They could also be used to optimize the starshade performance for particularly difficult observations.

## VI. Conclusions

In this paper a mathematical framework for understanding and analyzing starshade designs has been developed.

It was shown that "Offset Hypergaussians" provide an apodization that enables practical sized starshades to be built in support of direct observation of Earth-like planets. Formulae for the central depth of the shadow and its off-axis degradation have been derived.

It was shown how integration over radius can change the two-dimensional Fresnel integral into a one-dimensional edge integral in the case of binary optics. This is one approach to making computer algorithms fast enough to perform detailed analysis of the deep shadow.

It was shown how perturbation analysis of can be used to understand the basic tolerances of a starshade system and lead to simple scaling relations for such tolerances.

An analytic explanation for the behavior of the generalized apodization functions of VCK was developed and was used to explain why some of the shape tolerances for their generalized design can be much tighter than for the hypergaussian case.

In general, the analytic approach gives insight into the design and building of starshades that cannot be easily gained with computers alone. These results further support the sense of confidence that they can be built and flown.

Acknowledgements
I wish to thank the many colleagues who have worked with me in the lengthy process of coming to an understanding of how starshades function. In particular, I am indebted to J. Arenberg, T. Glassman, A. Lo, and R. Vanderbei. This work had been supported by grants from the NASA Institute for Advanced Concepts and NASA's Science Mission directorate. The engineering support provided by Northrop Grumman Aerospace Systems has been essential to the rapid development of starshades.